\documentclass[aps,pra,twocolumn,showpacs,amsmath,amssymb,floatfix,superscriptaddress]{revtex4-1}

\usepackage{graphicx}
\usepackage[bookmarks,bookmarksopen,bookmarksnumbered,colorlinks,linkcolor=red,linktocpage,citecolor=blue,urlcolor=cyan,pdfpagemode=UseOutline]{hyperref}

\def\sss{\scriptscriptstyle\rm}

\def\ben{\begin{equation}}
\def\een{\end{equation}}

\def\br{{\bf r}}

\def\ext{_{\rm ext}}
\def\ee{_{\rm ee}}
\def\xc{_{\sss XC}}
\def\Hx{_{\sss HX}}
\def\Hxc{_{\sss HXC}}
\def\x{_{\sss X}}
\def\c{_{\sss C}}
\def\H{_{\sss H}}

\def\wt{\mathtt{w}}
\def\calE{\mathcal{E}}
\def\calD{\mathcal{D}}
\def\calM{\mathcal{M}}
\def\bracew{{\{w\}}}
\def\txtKS{\text{KS}}
\def\txts{\text{s}}
\def\txtHxc{\text{Hxc}}
\def\txtHx{\text{Hx}}
\def\txtc{\text{c}}
\def\txtPT2{\text{PT2}}

\newcommand{\matelem}[3]{\left\langle #1 \left| #2 \right| #3 \right\rangle}

\newcommand{\intd}{\mathrm{d}}
\newcommand{\parref}[1]{(\ref{#1})}
\DeclareMathOperator{\tr}{tr}
\providecommand{\abs}[1]{\left|#1\right|}
\providecommand{\bra}[1]{\left< #1 \right|}
\providecommand{\ket}[1]{\left| #1 \right>}

\newcommand{\revision}[1]{{#1}}

\begin{document}

\title{Second-order perturbative correlation energy functional in the ensemble density-functional theory}
\author{Zeng-hui Yang}
\affiliation{Microsystem and Terahertz research center, China Academy of Engineering Physics, Chengdu, China 610200}
\affiliation{Institute of Electronic Engineering, China Academy of Engineering Physics, Mianyang, China 621000}
\date{\today}

\begin{abstract}
We derive the second-order approximation (PT2) to the ensemble correlation energy functional by applying the G\"{o}rling-Levy perturbation theory on the ensemble density-functional theory (EDFT). Its performance is checked by calculating excitation energies with the direct ensemble correction method in 1D model systems and 3D atoms using numerically exact Kohn-Sham orbitals and potentials. Comparing with the exchange-only approximation, the inclusion of the ensemble PT2 correlation improves the excitation energies in 1D model systems in most cases, including double excitations and charge-transfer excitations. However, the excitation energies for atoms are generally worse with PT2. We find that the failure of PT2 in atoms is due to the two contributions of an orbital-dependent functional to excitation energies being inconsistent in the calculations. We also analyze the convergence of PT2 excitation energies with respect to the number of unoccupied orbitals.
\end{abstract}

\maketitle

\section{Introduction}
\label{sec:intro}
The ensemble density functional theory (EDFT)\cite{T79,GOKb88,GOK88,OGKb88} is a formally exact excited-state extension to the highly successful density-functional theory (DFT)\cite{HK64,KS65,FNM03}. \revision{EDFT is shown to be promising in solving some of the difficult problems in the widely used time-dependent density-functional theory (TDDFT)\cite{RG84,C96,MMNG12,U12} such as the double excitation\cite{TH00,LKQM06,EGCM11,M16} charge-transfer excitation\cite{TAHR99,DWH03,DH04,M16,M17} and so on, and recently there is a renewed interest in EDFT\cite{F15,ADKF17,GP17,YPBU17,GKP18,SB18,DF19,GP19,LFM19,F20,G20,GSP20,LF20,MSFL20,FLNC21,GKP21,HH21}.}

One need to approximate the ensemble Hartree-exchange-correlation (Hxc) energy functional for practical calculations with EDFT. \revision{Unlike ground state DFT, it is more difficult to develop approximations as explicit density functionals, since ensemble functionals have an extra dependence on ensemble weights. Furthermore, the ensemble Hartree and exchange energies are not separated naturally\cite{GSP20} and an involved process is required to utilize ground-state experiences in approximations\cite{GKP21}, which is another difficulty in developing approximated functionals in EDFT. Only a few approximations are available in EDFT\cite{OGKb88,YPBU17,ADKF17,DMSF18,F20,LF20,MSFL20,GKP21}, and many are not built for general usage.}

\revision{One can avoid most of the difficulties associated with developing explicit density functionals by using orbital-dependent approximations.} The most simple one is the exact-exchange approximation known as symmetry-eigenstate Hartree-exchange\cite{PYTB14,YTPB14,YPBU17} or ensemble exact-exchange (EEXX)\cite{DMSF18,GKP18,SB18,SF18,DF19,SF20} in literature, the form of which is simply the definition of the ensemble Hartree-exchange energy. EEXX in general performs well, but the errors in the excitation energies can be large in some cases\cite{YPBU17,GKP18,SB18} due to the lack of correlation. An approximated ensemble correlation energy functional is needed to improve the accuracy.

\revision{In this paper, we derive an orbital-dependent second-order perturbative approximation (PT2) to the ensemble correlation energy functional by applying the G\"{o}rling-Levy perturbation theory (GLPT)\cite{GL93,GL94} on EDFT. Since EEXX can be obtained from the first order correction to the non-interacting Kohn-Sham (KS) ensemble energy, the PT2 correlation is the most simple natural extension to EEXX. Since the optimized effective potential (OEP)\cite{KLI92} in EDFT is more complicated than in DFT due to the dependence on ensemble weights, we test the PT2 correlation with the direct ensemble correction (DEC)\cite{YPBU17} instead of doing self-consistent ensemble Kohn-Sham (EKS) calculations.} We find that \revision{PT2} generally improves the accuracy of excitation energies in 1D model systems including double excitations and charge-transfer excitations. The performance of \revision{PT2} is unsatisfactory in 3D atoms, however, and we discuss the possible reasons. The convergence of \revision{PT2} excitation energies \revision{with respect to orbitals} is discussed as well.

\section{Theory}
\label{sec:theory}
\subsection{Background}
\label{sec:theory:background}
We use atomic units [$e=\hbar=m_e=1/(4\pi\epsilon_0)=1$] unless otherwise specified. The interacting system is described by the Sch\"{o}dinger equation:
\ben
\hat{H}\ket{\Psi_{ik}}=\calE_i\ket{\Psi_{ik}},
\label{eqn:theory:general:schrodinger}
\een
where the Hamiltonian $\hat{H}$ is $\hat{H}=\hat{T}+\hat{V}\ext+\hat{V}\ee$ with $\hat{T}$, $\hat{V}\ext$ and $\hat{V}\ee$ being the kinetic, external potential and electron-electron interaction potential operators, and $i$ denotes a set of degenerate states (`multiplet') and $k$ an individual state in the multiplet. The multiplets are ordered by energy \revision{with} $i=0$ \revision{being} the ground state. The ordering of wavefunctions within a multiplet is arbitrary but fixed. An ensemble in EDFT contains consecutive multiplets from the ground state to a highest-energy multiplet $I$. Each state in multiplet $i$ is assigned a weight $w_i$ satisfying $w_i\ge w_j$ for $i<j$.\cite{GOKb88} We require $\sum_{i=0}^I g_i w_i=1$ for simplicity, where $g_i$ is the degeneracy of multiplet $i$. We denote all weights as $\bracew$ in the following.

The ensemble density and energy are defined as
\ben
n_\bracew(\br)=\tr\{\hat{D}_\bracew\hat{n}(\br)\}=\sum_{i=0}^I w_i\sum_{k=1}^{g_i} n_{ik}(\br),
\label{eqn:theory:general:ensden}
\een
and
\ben
E_\bracew=\tr\{\hat{D}_\bracew\hat{H}\}=\sum_{i=0}^I w_i\sum_{k=1}^{g_i} \calE_i,
\label{eqn:theory:general:enseng}
\een
where the ensemble density matrix $\hat{D}_\bracew$ is
\ben
\hat{D}_\bracew=\sum_{i=0}^I w_i\sum_{k=1}^{g_i}\ket{\Psi_{ik}}\bra{\Psi_{ik}},
\label{eqn:theory:general:ensDM}
\een
and $n_{ik}(\br)=\matelem{\Psi_{ik}}{\hat{n}(\br)}{\Psi_{ik}}$.

\revision{EDFT proves the one-to-one correspondence betweeen $n_\bracew$ and the external potential given $\hat{V}\ee$\cite{GOKb88}, so $E_\bracew$ is a functional of $n_\bracew$}. One can define a non-interacting \revision{EKS} system with the same $n_\bracew$ as the interacting system, with the EKS orbitals satisfying
\ben
\left\{-\frac{1}{2}\nabla^2_\br+v_{\txts,\bracew}[n_\bracew](\br)\right\}\phi_{\mu,\bracew}(\br)=\epsilon_{\mu,\bracew}\phi_{\mu,\bracew}(\br).
\label{eqn:theory:general:ensKSeqn}
\een
The EKS density matrix $\hat{D}_{\txts,\bracew}$ is
\ben
\hat{D}_{\txts,\bracew}=\sum_{i=0}^{I} w_i\sum_{k=1}^{g_i} \ket{\Phi_{ik,\bracew}}\bra{\Phi_{ik,\bracew}},
\label{eqn:theory:general:ensKSDM}
\een
where $\ket{\Phi_{ik,\bracew}}$ is an EKS wavefunction corresponding to the $\ket{\Psi_{ik}}$ state of the interacting system. The existence of an adiabatic connection\cite{PY89,FNM03} between the \revision{interacting and EKS systems} is assumed. $n^\txtKS_\bracew(\br)=\tr\{\hat{D}_{\txts,\bracew}\hat{n}(\br)\}=n_\bracew(\br)$ \revision{if} $v_{\txts,\bracew}$ is exact.

Unlike in ground-state DFT, $\ket{\Phi_{ik,\bracew}}$ is not necessarily a single Slater determinant of EKS orbitals since the EKS degenercies are in general greater than or equal to the corresponding interacting ones.\cite{PYTB14} We require $\Phi_{ik,\bracew}$ to have the same spatial and spin symmetries as $\Psi_{ik}$\cite{PYTB14} to distinguish the otherwise degenerate EKS states by linearly combine the EKS Slater determinants:
\ben
\ket{\Phi_{ik,\bracew}}=\sum_{p=1}^{\tilde{g}_{\tilde{i}}}C_{ikp}\ket{\tilde{\Phi}_{\tilde{i}p,\bracew}},
\label{eqn:DECSEHX:symwf}
\een
where $\tilde{i}$ and $\tilde{g}_{\tilde{i}}$ denote the EKS multiplet and its degeneracy corresponding to interacting multiplet $i$, $\tilde{\Phi}_{\tilde{i}p,w}$ denotes the EKS Slater determinant $p$ in EKS multiplet $\tilde{i}$, and $C_{ikp}$ is the linear combination coefficient.

$E_\bracew$ can be decomposed as
\ben
\begin{split}
E_\bracew[n]=&T_{\txts,\bracew}[n]+V\ext[n]+E_{\txtHx,\bracew}[n]+E_{\txtc,\bracew}[n]\\
=&\tr\{\hat{D}_{\txts,\bracew}\hat{T}\}+\int\intd^3r\,n(\br)v\ext(\br)\\
&+\tr\{\hat{D}_{\txts,\bracew}\hat{V}\ee\}+E_{\txtc,\bracew}[n].
\end{split}
\label{eqn:theory:background:Edecomp}
\een
The ensemble Hartree-exchange energy is
\ben
E_{\txtHx,\bracew}=\tr\{\hat{D}_{\txts,\bracew}\revision{\hat{V}\ee}\}=\sum_{i=0}^I w_i\sum_{k=1}^{g_i} \matelem{\Phi_{ik,\bracew}}{\hat{V}\ee}{\Phi_{ik,\bracew}},
\label{eqn:DECSEHX:EHx}
\een
which can be evaluated using the Slater-Condon rules\cite{L14}. The ensemble correlation energy $E_{\txtc,\bracew}$ is defined by Eq. \parref{eqn:theory:background:Edecomp} and must be approximated in practice. The EKS potential $v_{\txts,\bracew}(\br)$ is determined \revision{variationally by $\delta E_\bracew[n]/\delta n(\br)=0$ as}
\ben
\begin{split}
v_{\txts,\bracew}[n](\br)&=v\ext(\br)+v_{\txtHx,\bracew}[n](\br)+v_{\txtc,\bracew}[n](\br)\\
&=v\ext(\br)+\frac{\delta E_{\txtHx,\bracew}[n]}{\delta n(\br)}+\frac{\delta E_{\txtc,\bracew}[n]}{\delta n(\br)}.
\end{split}
\label{eqn:theory:background:vdecomp}
\een

EDFT calculates excitation energies either by subtracting ensemble energies or by taking derivative of $E_\bracew$ with respect to the weights\cite{GOK88,DF19}, and both requires solving Eq. \parref{eqn:theory:general:ensKSeqn} self-consistently. The DEC method of our previous work\cite{YPBU17} can be used to calculate excitation energies without extra self-consistent calculations other than the ground-state one. \revision{The DEC method uses a special type of ensemble defined as}
\ben
w_i(\wt)=\left\{\begin{array}{cl}
\frac{1-\wt(M_I-g_0)}{g_0} & i=0,\\
\wt & i\ne0,
\end{array}\right.
\label{eqn:DECSEHX:GOKII}
\een
where $M_I=\sum_i g_i$ is the total number of states in the ensemble, and $\wt\in[0,1/M_I]$ is the weight parameter. The excitation energy $\omega_I=\calE_I-\calE_0$ can be written as a correction to the ground-state KS excitation energy:
\ben
\revision{
\omega_I=\omega_I^\txtKS+\frac{1}{g_I}\left.\frac{d}{dw}(E_{\txtHxc,I,\wt}-E_{\txtHxc,I-1,\wt})\right|_{\wt=0}.
}
\label{eqn:DECSEHX:DEC}
\een
\revision{The ensemble density reduces to the ground-state density at $\wt=0$}, so Eq. \parref{eqn:DECSEHX:DEC} only requires a self-consistent ground-state KS calculation.

\subsection{GLPT of EDFT}
\label{sec:theory:GLPT}
\revision{GLPT\cite{GL93,GL94} is developed for ground-state DFT as a systematic approach for developing orbital-dependent xc functionals. The perturbation is along the adiabatic connection with the KS Hamiltonian as the zeroth order and density fixed. The perturbation is the difference between the interacting and KS Hamiltonians $\hat{V}\ee-\hat{V}\Hxc$, so one can extract approximations to $E\xc$ and $v\xc$ from the perturbative corrections. GLPT differs from the Rayleigh-Schr\"{o}dinger perturbation theory since the perturbation is dependent on the coupling strength, and perturbative corrections to the density vanish at all orders. GLPT yields the exact exchange (EXX) approximation which is widely used in both DFT and TDDFT.}

\revision{We apply GLPT on EDFT in this paper. GLPT has to be applied along a different adiabatic connection with the ensemble density fixed, with the EKS Hamiltonian as the zeroth-order Hamiltonian.} The Schr\"{o}dinger equation with the electron-electron interaction scaled by the coupling constant $\lambda\in[0,1]$ is
\ben
\revision{(\hat{T}+\hat{V}_{\lambda,\bracew}+\lambda\hat{V}\ee)}\ket{\Phi_{ik,\lambda,\bracew}}=\calE_{i,\lambda,\bracew}\ket{\Phi_{ik,\lambda,\bracew}},
\label{eqn:couplingconst:schrodingereqn}
\een
where \revision{$\hat{V}_{\lambda,\bracew}=\sum_i v_{\lambda,\bracew}(\br_i)$}, and the existence of $v_{\lambda,\bracew}$ is assumed. The \revision{scaled ensemble energy is}
\ben
\begin{split}
E_{\lambda,\bracew}&=\tr\{\hat{D}_{\lambda,\bracew}\hat{H}_\lambda\}=\sum_{i=0}^I w_i g_i \calE_{i,\lambda,\bracew}\\
&=T_{\txts,\lambda,\bracew}+\int\intd^3r\,n_{\lambda,\bracew}(\br)\left[v\ext(\br)+v_{\txtHxc,\lambda,\bracew}(\br)\right],
\end{split}
\label{eqn:theory:GLPT:enseng}
\een
where $\hat{D}_{\lambda,\bracew}=\sum_{i=0}^I w_i\sum_{k=1}^{g_i}\ket{\Phi_{ik,\lambda,\bracew}}\bra{\Phi_{ik,\lambda,\bracew}}$. \revision{Treating $\lambda\hat{V}\ee+\hat{V}_{\lambda,\bracew}-\hat{V}_{\txts,\bracew}$ as the perturbation and expanding $v_{\lambda,\bracew}(\br)$ as $\sum_{p=0}^\infty\lambda^p v^{(p)}_\bracew(\br)$, we obtain the formula for the ensemble PT2 correlation energy:}
\ben
E^\txtPT2_{\txtc,\bracew}=\sum_{i=0}^I w_i\sum_{k=1}^{g_i}\matelem{\Phi_{ik,\bracew}}{\hat{V}\ee+\hat{V}^{(1)}_\bracew}{\Phi^{(1)}_{ik,\bracew}}.
\label{eqn:theory:GLPT:EcPT2}
\een
\revision{Details of the derivations is available in Appendix \ref{app:GLPT}.}

Degenerate perturbation theory\cite{D61,L14} requires that the zeroth-order wavefunctions of an EKS multiplet diagonalizes $\hat{H}'$, and this is satisfied by the EKS wavefunctions in Eq. \parref{eqn:DECSEHX:symwf} since they are eigenfunctions of the spatial and spin symmetry operators that commute with $\hat{H}$. $\ket{\Phi^{(1)}_{ik,\bracew}}$ in Eq. \parref{eqn:theory:GLPT:EcPT2} is then
\ben
\ket{\Phi^{(1)}_{ik,\bracew}}=\sum_{\tilde{j}\ne\tilde{i}}^\infty\sum_{q=1}^{\tilde{g}_{\tilde{j}}} \frac{\matelem{\tilde{\Phi}^{(0)}_{\tilde{j}q,\bracew}}{\hat{V}\ee+\hat{V}^{(1)}_\bracew}{\Phi^{(0)}_{ik,\bracew}}}{\calE^\txtKS_{i,\bracew}-\calE^\txtKS_{j,\bracew}} \ket{\tilde{\Phi}^{(0)}_{\tilde{j}q,\bracew}},
\label{eqn:PT2:phi1}
\een
where $\calE^\txtKS_{i,\bracew}$ is the EKS energy of EKS multiplet $\tilde{i}$.

The OEP equations for the \revision{perturbative corrections of the potential} can be derived \revision{by requiring the perturbative corrections of the density to vanish\cite{GL93}}. However, these equations involve weighted sums over states and are more difficult to solve than their counterparts in ground-state DFT. Approximations to $v^{(1)}_\bracew$ can be derived and are provided in the supplemental material\cite{supplemental}. While approximations to $v^{(2)}_\bracew$ can be derived similarly in principle, it is more cumbersome as it would contain functional derivatives of $v^{(1)}_\bracew$, so self-consistent EKS calculation with ensemble PT2 correlation can be unhandy. We also encounter divergences related to the ensemble derivative discontinuity\cite{L95} in the OEP calculations. We therefore choose to use the DEC method to assess the performance of the ensemble PT2 correlation in this paper to avoid solving the EKS equation self-consistently.

\subsection{Ensemble PT2 correlation in the DEC method}
\label{sec:theory:DECPT2}
The orbital-dependent Eqs. \parref{eqn:DECSEHX:EHx} and \parref{eqn:theory:GLPT:EcPT2} are implicit density functionals. \revision{In the DEC method, these equations represent $E_{\txtHx,\wt}[n^\txtKS_\wt]$ and $E^\txtPT2_{\txtc,\wt}[n^\txtKS_\wt]$ instead of $E_{\txtHx,\wt}[n]$ and $E^\txtPT2_{\txtc,\wt}[n]$ required by Eq. \parref{eqn:DECSEHX:DEC}}. Since the $\wt$-dependences of the functional and of the EKS density are inseparable in \revision{these equations}, \revision{we calculate the $\wt$-derivatives in Eq. \parref{eqn:DECSEHX:DEC} by}\cite{YPBU17}
\ben
\revision{
\left.\frac{d E_{\txtHxc,I,\wt}}{d\wt}\right|_{\wt=0}=\calD\left(E_{\txtHxc,I,\wt}\right)-\int\intd^3r\,v\Hxc(\br)\calD\left(n^\txtKS_{I,\wt}\right),
}
\label{eqn:DECSEHX:derivdetail}
\een
where $v\Hxc$ is the ground-state Hxc potential, and the shorthand $\calD$ means
\revision{
\begin{align}
\calD\left(E_{\txtHxc,I,\wt}\right)&=\left.\frac{\partial E_{\txtHxc,I,\wt}[n^\txtKS_{I,\wt}[\{\phi_\mu\}]]}{\partial\wt}\right|_{\substack{\wt=0\\ \{\phi_\mu\}=\{\phi^\txtKS_{\mu}\}}},\\
\calD\left(n^\txtKS_{I,\wt}\right)&=\left.\frac{\partial n^\txtKS_{I,\wt}[\{\phi_\mu\}](\br)}{\partial\wt}\right|_{\substack{\wt=0\\ \{\phi_\mu\}=\{\phi^\txtKS_{\mu}\}}},
\end{align}
}
with $\{\phi_\mu\}$ and $\{\phi^\txtKS_{\mu}\}$ being a complete set of one-electron orbitals and the ground-state KS orbitals respectively, and $n^\txtKS_{I,\wt}[\{\phi_\mu\}]$ being the EKS density of the ensemble \revision{defined by Eq. \parref{eqn:DECSEHX:GOKII}} with $I$ being the highest energy multiplet constructed with orbitals $\{\phi_\mu\}$.

The orbitals are held fixed for the derivatives in Eq. \parref{eqn:DECSEHX:derivdetail} to avoid \revision{the otherwise required self-consistent EKS calculations}, since the $\wt$-dependence of \revision{$\ket{\tilde{\Phi}_{\tilde{i}p,\wt}}$ and $E^\txtKS_{i,\wt}$} originates from that of the EKS orbitals. \revision{$\calD(E_{\txtHxc,I,\wt})$ is then the direct differentiation of Eqs. \parref{eqn:DECSEHX:EHx} and \parref{eqn:theory:GLPT:EcPT2}. Its PT2 part is}
\begin{multline}
\calD\left(E^\txtPT2_{\txtc,I,\wt}\right)=\sum_{i=0}^I \calD(w_i)\sum_{k=1}^{g_i}\sum_{\tilde{j}\ne\tilde{i}}^\infty\sum_{q=1}^{\tilde{g}_{\tilde{j}}}\frac{\abs{\calM_{\tilde{j}q}^{ik}\left(\hat{V}\ee-\hat{V}\Hx\right)}^2} {\calE^\txtKS_{i,\wt}-\calE^\txtKS_{j,\wt}}\\
+\frac{1}{g_0}\sum_{k=1}^{g_0} \sum_{\tilde{j}\ne\tilde{0}}^\infty\sum_{q=1}^{\tilde{g}_{\tilde{j}}}\frac{
\left\{\calM_{\tilde{j}q}^{ik}\left(\hat{V}\ee-\hat{V}\Hx\right)^*\calM_{\tilde{j}q}^{0k}\left[-\calD\left(\hat{V}_{\txtHx,\wt}\right)\right]+\text{c.c.}\right\}
}{\calE^\txtKS_{0}-\calE^\txtKS_{j}}
\label{eqn:PT2:derivPT2work}
\end{multline}
where $\calD(w_i)=d w_i(\wt)/d\wt|_{\wt=0}$, $\calD(\hat{V}_{\txtHx,\wt})=\partial\hat{V}_{\txtHx,\wt}[n_{\txts,\wt}[\{\phi_\mu\}]]/\partial \wt|_{\{\phi_\mu\}=\{\phi^\txtKS_\mu\},\wt=0}$, and the notation $\calM$ means
\begin{align}
\calM_{\tilde{j}q}^{ik}\left(\hat{V}\ee-\hat{V}\Hx\right)&=\matelem{\tilde{\Phi}_{\tilde{j}q}}{ \hat{V}\ee-\hat{V}\Hx}{\Phi_{ik}},\\
\calM_{\tilde{j}q}^{0k}\left[-\calD\left(\hat{V}_{\txtHx,\wt}\right)\right]&=\matelem{\tilde{\Phi}_{\tilde{j}q}}{-\calD\left(\hat{V}_{\txtHx,\wt}\right)}{\Phi_{ik}}.
\end{align}

\revision{We ignore the second term of Eq. \parref{eqn:PT2:derivPT2work} in the calculations in this paper since it is usually small (\revision{Appendix \ref{sec:OEP:justification}}). This approximation allows us to greatly simplify the calculation by using a bi-ensemble comprised of the ground state and the $I$-th multiplet, which is equivalent to the full ensemble due to the functional form of the first term of Eq. \parref{eqn:PT2:derivPT2work}.\cite{YPBU17} Eq. \parref{eqn:PT2:derivPT2work} apparently indicates that the number of orbitals required for convergence determines the scaling of the computational cost.}

\section{Results and discussion}
\label{sec:results}
We carry out \revision{DEC} calculations \revision{with EEXX and EEXX+PT2 functionals} on 1D model systems for double excitation and charge-transfer excitation and on 3D He and Be atoms. TDDFT results are reported as well for comparison. \revision{We use the numerically exact KS ground state and the exact $v\Hx$ and $v\Hxc$ in the calculations. For EEXX calculations, the errors in excitation energies would exclusively reflect the performances of the method; for EEXX+PT2, however, the results contain higher-order terms due to the exact $v\c$, which is used to avoid solving for the OEP $v\c^\txtPT2$.}

\revision{The} orbital-dependent ensemble Hxc energy functional contributes to the excitation energy in two different ways corresponding to the two terms of Eq. \parref{eqn:DECSEHX:derivdetail}, one from the $\wt$-dependence of the orbital-dependent functional, and another one from the Hxc potential. This is relevant to self-consistent EKS calculations as well since the excitation energies also depend on $\partial E_{\txtHxc,I,\wt}/\partial\wt$\cite{GOK88}. \revision{For the EEXX functional, the two terms of Eq. \parref{eqn:DECSEHX:derivdetail} are found to have the same orders of magnitude\cite{YPBU17}, but this is not true for the correlation part. We demonstrate this by reporting excitation energies of EEXX+$E\c^\txtPT2$ and EEXX+$v\c$, where $E\c^\txtPT2$ and $v\c$ denote the correlation contribution to the two terms of Eq. \parref{eqn:DECSEHX:derivdetail} respectively. The EEXX+PT2 result can be written as EEXX+$E\c^\txtPT2$+$v\c$ in this notation.}

In ground-state DFT, a common approximation to the PT2 correlation energy is to neglect the `singly-excited' matrix elements where the Slater determinants of the bra and ket differ by one orbital (such as in the B2PLYP\cite{G06} double hybrid functional). We test the performance of this approximation as well and denote it as \revision{PT2(no single)}.

\subsection{1D model systems}
\label{sec:results:1D}
We solve the interacting Schr\"{o}dinger equation for 1D model systems by direct diagonalization on a grid. All 1D model systems have two electrons so that the exact $v\x(x)$ can be obtained as $-v\H(x)/2$. The second term of Eq. \parref{eqn:PT2:derivPT2work} vanishes for two-electron systems, so the 1D PT2 calculations are not approximated. Numerically exact \revision{KS ground state is} obtained by inverting the ground-state KS equation\cite{YTPB14}.

We use an 1D Hooke's atom as the model system for double excitation \cite{MZCB04,YPBU17}. The system has $v\ext(x)=x^2/2$ and contact interaction $v\ee(x,x')=0.2\delta(x-x')$, where $\delta$ is the Dirac $\delta$ function. The doubly excited states in this system are close to singly excited states due to the weak electron-electron interaction. We use a large $20001\times20001$ grid with grid-point spacing $0.001$ and $x\in[-10,10]$ to ensure that the effect of the grid boundary on all orbitals is negligible. Table \ref{table:results:1D:double} lists the errors in the excitation energies.

\begin{table}[htbp]
\footnotesize
\begin{tabular}{cccccccc}
\hline\hline
 & \multicolumn{7}{c}{$\Delta\omega_I=\omega_I-\omega_I^\text{exact}$ (mH)}\\
\cline{2-8}
 & \multicolumn{5}{c}{DEC} & TDDFT & TDDFT\\
\cline{2-6}
$I$ & EEXX & \revision{+$E\c^\txtPT2$} & \revision{+$v\c$} & \revision{+}PT2 & \revision{$\substack{\text{+PT2}\\\text{(no single)}}$} & AEXX & DSPA\\
1(1,2) & 1.389 & \revision{2.240} & \revision{1.350} & 2.201 & 2.401 & 1.041 & 1.400\\
2(2,2) & 17.24 & \revision{4.565} & \revision{17.16} & 4.487 & 5.001 & N/A & -1.900\\
3(1,3) & -16.65 & \revision{-1.929} & \revision{-18.27} & -3.550 & -3.554 & -16.94 & 2.200\\
4(2,3) & 28.34 & \revision{19.85} & \revision{26.68} & 18.19 & 18.15 & N/A & -1.800\\
5(1,4) & -26.60 & \revision{-15.78} & \revision{-28.40} & -17.58 & -17.05 & -26.55 & 1.600\\
\hline\hline
\end{tabular}
\caption{Errors in the first 5 singlet excitation energies of the 1D Hooke's atom\revision{, where $I=2$ and $I=4$ are doubly excited states. +$E\c^\txtPT2$ is short for EEXX+$E\c^\txtPT2$, and the same goes for +$v\c$, +PT2 and +PT2(no single).} The occupied KS orbitals \revision{of the excited states} are shown in parentheses, and the ground state is $(1,1)$. AEXX denotes the adiabatic exact-exchange kernel in TDDFT\cite{MZCB04}. All calculations use 10 orbitals. DSPA is the dressed TDDFT single-pole approximation of Ref. \cite{MZCB04} for comparison.}
\label{table:results:1D:double}
\end{table}

\revision{$\Delta\omega_I^\text{EEXX}$ shows the error in excitation energies due to the lack of correlation. With PT2 correlation, EEXX+PT2 in Table \ref{table:results:1D:double} improves the accuracy of excitation energies }in most cases, including both doubly-excited states. Due to the perturbative nature of the ensemble PT2 correlation, however, improvement in accuracy is not guaranteed, as seen in the first excited state.

\revision{The correlation contribution of the two terms of Eq. \parref{eqn:DECSEHX:derivdetail} can be obtained from Table \ref{table:results:1D:double} by subtracting column 1(EEXX) from columns 2(EEXX+$E\c^\txtPT2$) and 3(EEXX+$v\c$), which shows that the correlation effect in the 1D Hooke's atom is dominated by the first term of Eq. \parref{eqn:DECSEHX:derivdetail}. This dominance allows us to estimate the PT2 errors in excitation energies with this first term alone while ignoring the effect of using exact $v\c$: comparing $\omega_I^\text{exact}-\omega_I^{\text{EEXX}+v\c}$ (the effect of this term of the exact functional) and $\omega_I^{\text{EEXX}+E\c^\txtPT2}-\omega_I^\text{EEXX}$ (that of the PT2 approximation) yields $163\%$, $26\%$, $19\%$, $19\%$, $62\%$ for the 5 states in Table \ref{table:results:1D:double}. These large errors for PT2 indicate that higher-order terms are probably needed to properly describe the ensemble correlation energy.}

We use a system with two potential wells in a box as the model system for charge-transfer excitation. The external potential is
\ben
v\ext(x)=\left\{\begin{array}{ll}
\infty & x\in(-\infty,0]\cup[6.5,\infty),\\
0 & x\in(0,1)\cup(5,6.5),\\
20 & x\in[1,5].
\end{array}\right.
\label{eqn:results:1D:CT:vext}
\een
and the electron-electron interaction is the soft-Coulomb potential
\ben
v\ee(x,x')=\frac{1}{\sqrt{(x-x')^2+\alpha}}.
\label{eqn:results:1D:veesoftcoulomb}
\een
We set $\alpha=1$ in order to carry out adiabatic TDDFT calculations with the 1D local-density approximation (LDA)\cite{HFCV11} kernel for comparison. We orbtain the numerically exact KS system on a $1301\times1301$ grid with grid-point spacing $0.005$ and $x\in[0,6.5]$. Excitations to the first (triplet) and second (singlet) excited states are charge-transfer excitations, both of which correspond to one electron transferring from the right potential well to the left one. The distance between the two wells lead to negligible overlap between the involved KS orbitals, so the TDDFT couplings between the initial and final states vanish for local or semi-local approximated xc kernels. The corresponding TDDFT excitation energies would be very close to the KS ones.

\begin{table}[htbp]
\begin{tabular}{ccccccc}
\hline\hline
\multicolumn{7}{c}{$\Delta\omega_1=\omega_1-\omega_1^\text{exact}$ (mH)}\\
\hline
$KS$ & \multicolumn{5}{c}{DEC} & TDDFT\\
\cline{2-6}
(mH) & EEXX & \revision{+$E\c^\txtPT2$} & \revision{+$v\c$} & \revision{+}PT2 & \revision{$\substack{\text{+PT2}\\\text{(no single)}}$} & ALDA\\
-53.38 & -53.38 & \revision{-53.18} & \revision{-0.1011} & 0.1027 & 0.2205 & -53.38\\
\hline\hline
\end{tabular}
\caption{Errors in the first ($I=1$) charge-transfer excitation energy of the 1D model system described in Eq. \parref{eqn:results:1D:CT:vext}. ALDA denotes the adiabatic LDA xc kernel. 7 orbitals are used in both \revision{EEXX+PT2} and in TDDFT.}
\label{table:results:1D:CT}
\end{table}

We list the error in the KS excitation energy in Table \ref{table:results:1D:CT}\revision{. Since the singlet-triplet splitting between the first two states is very small, Table \ref{table:results:1D:CT} only lists the calculation result for the first excited state. The} correlation effect dominates the correction to $\omega_I^\txtKS$ for the charge-transfer box \revision{as well}. Both DEC/EEXX and TDDFT/ALDA fail to correct the KS excitation energy due to non-overlapping orbitals leading to vanishing matrix elements. Unlike the 1D Hooke's atom, the correlation contribution from the $\wt$-dependence of the orbital-dependent $E^\text{PT2}_{\txtc,\wt}$ \revision{[first term of Eq. \parref{eqn:DECSEHX:derivdetail}]} only changes the result slightly due to either vanishing matrix elements or large energy differences in Eq. \parref{eqn:PT2:derivPT2work}. The contribution from $v\c$ dominates the correlation effect. This suggests that the two correlation contributions in Eq. \parref{eqn:DECSEHX:derivdetail} can be regarded as representing the local and the non-local correlation effects, respectively.

We also \revision{calculate} the excitation energies of an 1D flat box, where both terms of Eq. \parref{eqn:DECSEHX:derivdetail} contribute significantly to the correlation effect \revision{(Appendix \ref{app:1dbox})}. \revision{We find that the inclusion of ensemble PT2 correlation improves the exchange-only EEXX excitation energies for most of the states of the tested 1D systems.} However, the PT2 correction to EEXX is not always in the correct direction. The magnitude of improvement also varies a lot for different states.

\subsection{3D atoms}
\label{sec:results:3D}
We \revision{carry out DEC calculations} on He and Be atoms using numerically exact KS potentials\cite{UG93,UG94}, and compare \revision{the excitation energies} to the experimental values\cite{NIST_ASD}. Since we solve the ground-state KS orbitals on a radial grid, the higher-energy orbitals are not well represented \revision{due to tails being} truncated at the grid boundary. This leads to the ordering of higher-energy orbitals being different from that of the exact KS system (see He PT2 convergece curve in supplemental material\cite{supplemental}). Nevertheless, this should not affect the PT2 results since the orbitals form a complete basis set with respect to the grid, and convergence is achieved fairly quickly for PT2 (Sec. \ref{sec:results:discussion}).

\begin{table}[htbp]
\begin{tabular}{ccccccc}
\hline\hline
 & \multicolumn{6}{c}{$\revision{\Delta\omega_I=}\omega_I-\omega_I^\text{exact}$ (mH)}\\
\cline{2-7}
 & \multicolumn{5}{c}{DEC} & TDDFT\\
\cline{2-6}
$I$ & \revision{EEXX} & \revision{+$E\c^\txtPT2$} & \revision{+$v\c$} & \revision{+}PT2 & \revision{$\substack{\text{+PT2}\\\text{(no single)}}$} & ALDA\\
\hline
\multicolumn{7}{c}{He atom}\\
\hline
$^3$S(1s2s) & -4.996 & \revision{-23.93} & \revision{9.092} & -9.841 & 9.318 & -8.724\\
$^1$S(1s2s) & 11.25 & \revision{2.476} & \revision{25.34} & 16.56 & 23.16 & 10.61\\
$^3$P(1s2p) & -0.9324 & \revision{-22.22} & \revision{12.91} & -8.376 & 13.48 & -16.41\\
$^1$P(1s2p) & 5.355 & \revision{-21.59} & \revision{19.20} & -7.748 & 19.45 & -3.247\\
$^3$S(1s3s) & -1.075 & \revision{-24.14} & \revision{13.48} & -9.581 & 14.65 & -0.3482\\
$^1$S(1s3s) & 2.574 & \revision{-17.12} & \revision{17.13} & -2.566 & 18.57 & 4.217\\
$^3$P(1s3p) & -0.1077 & \revision{-24.22} & \revision{14.33} & -9.776 & 15.21 & -2.276\\
$^3$D(1s3d) & 0.3006 & \revision{-24.97} & \revision{15.02} & -10.24 & 15.92 & -1.036\\
$^1$D(1s3d) & 0.3715 & \revision{-25.09} & \revision{15.09} & -10.37 & 15.95 & -0.6598\\
$^1$P(1s3p) & 1.712 & \revision{-23.70} & \revision{16.15} & -9.257 & 17.17 & -0.0629\\
$^3$S(1s4s) & -0.2530 & \revision{-24.47} & \revision{14.39} & -9.822 & 15.50 & 0.2161\\
$^1$S(1s4s) & 1.140 & \revision{-21.70} & \revision{15.79} & -7.049 & 17.20 & 2.354\\
\hline
\multicolumn{7}{c}{Be atom}\\
\hline
$^3$P(1s$^2$2s2p) & -38.06 & \revision{15.46} & \revision{-38.76} & 14.76 & 17.59 & -46.89\\
$^1$P(1s$^2$2s2p) & 4.067 & \revision{52.91} & \revision{3.368} & 52.22 & 52.11 & 6.473\\
$^3$S(1s$^2$2s3s) & -4.407 & \revision{42.58} & \revision{-36.84} & 10.14 & 17.30 & -13.64\\
$^1$S(1s$^2$2s3s) & 6.179 & \revision{41.58} & \revision{-26.26} & 9.145 & 11.86 & 1.836\\
$^1$D(1s$^2$2p$^2$) & 10.77 & \revision{17.78} & \revision{9.374} & 16.38 & 17.55 & N/A\\
$^3$P(1s$^2$2s3p) & -4.778 & \revision{45.19} & \revision{-37.64} & 12.33 & 20.00 & -13.32\\
$^3$P(1s$^2$2p$^2$) & -35.49 & \revision{19.19} & \revision{-36.88} & 17.79 & 20.51 & N/A\\
\hline\hline
\end{tabular}
\caption{Errors in the excitation energies of the He and Be atoms. The final KS states of excitations are written in parentheses. 15 orbitals and 20 orbitals are used in calculations for He and Be respectively. The Tamm-Dancoff approximation (TDA)\cite{HH99} is used in TDDFT/ALDA calculations for Be due to numerical problems without TDA.}
\label{table:results:3D}
\end{table}

Table \ref{table:results:3D} lists the \revision{errors in the} excitation energies for He and Be. The performance of the ensemble PT2 correlation energy functional is disappointing in atoms. Unlike 1D systems, \revision{EEXX+PT2} excitation energies are in general worse than those of \revision{EEXX} despite a few exceptions. This includes the He atom as well, which is a two-electron system as the 1D model systems. The ground-state PT2 correlation is also found to be problematic when used directly\cite{MWY05}, and the failure is attributed to the algebraic structure of KS orbitals. However, this is unlikely the reason for the failure seen in Table \ref{table:results:3D} since the DEC method does not involve self-consistent ensemble PT2 calculations.

The reason for larger error of \revision{EEXX+PT2} might be related to the two terms of Eq. \parref{eqn:DECSEHX:derivdetail} being incompatible. For \revision{EEXX}, these two terms are close in magnitude and have different signs\cite{YPBU17}. Due to their cancellation, the resulting corrections to KS excitation energies have much smaller magnitudes. If these two terms are incompatible, the error can be large since Eq. \parref{eqn:DECSEHX:derivdetail} subtracts two large numbers, which can be seen from the \revision{EEXX+$v\c$} results. This might be the case for \revision{EEXX+PT2} shown in Table \ref{table:results:3D}, where we use the exact $v\c$ in Eq. \parref{eqn:DECSEHX:derivdetail} instead of the OEP $v\c^\txtPT2$. As an example to demonstrate the effect of incompatible $E_{\txtc,\wt}$ and $v\c$, using the LDA correlation potential\cite{PW92} in \revision{an EEXX} calculation of the first excitation energy of He yields an error of -55.98 mH, which is much greater than the one in Table \ref{table:results:3D}. \revision{Since the correlation is dominated by one of the two terms for} the 1D Hooke's atom and charge-transfer box, the error due to this incompatibility is small \revision{(but the error due to PT2 itself can still be large)}, so the overall accuracy of \revision{EEXX+PT2} only \revision{depends} on the dominant term. This problem is a disadvantage of orbital-dependent Hxc energy functionals and does not affect explicit density functionals. This failure also indicates that the OEP $v\c^\txtPT2$ is not close to the exact $v\c$ for atoms, which agrees with the large relative errors due to PT2 in the 1D Hooke's atom, showing that higher-order terms are probably needed for correlation.

\revision{EEXX+PT2(no single)} results for the 1D Hooke's atom and 1D charge-transfer box are close to the DEC/EEXX+PT2 results, but the error is larger for atoms. Despite this, the error introduced by this approximation is still small comparing with the excitation energies (see supplemental material\cite{supplemental} and Ref. \cite{NIST_ASD}), suggesting that the ground-state approximation of neglecting singly-excited matrix elements is also applicable in EDFT.

\subsection{Convergence for PT2}
\label{sec:results:discussion}
\begin{figure}[htbp]
\includegraphics[width=\columnwidth]{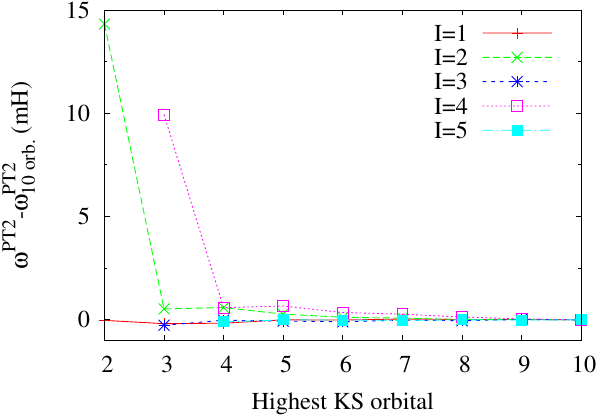}
\caption{Convergence of \revision{EEXX+PT2} excitation energies of the 1D Hooke's atom with respect to the number of KS orbitals. The excitation energies calculated with 10 KS orbitals are used as references. The differences between the last two points of each curve are $0.02$ mH, $0.02$ mH, $0.02$ mH, $0.05$ mH, and $0.006$ mH respectively.}
\label{fig:results:discussion:conv1dHooke}
\end{figure}

\revision{Although the} $\tilde{j}$- and $q$-sums in Eq. \parref{eqn:PT2:derivPT2work} sums over all KS states\revision{, these sums can be truncated by only including a finite number of KS orbitals, since the contributions from higher-energy orbitals vanish due to the KS energy differences in the denominators}. We check the convergence of the \revision{EEXX+PT2} excitation energies with respect to the number of KS orbitals, which are plotted in Fig. \ref{fig:results:discussion:conv1dHooke} for the 1D Hooke's atom.

Fig. \ref{fig:results:discussion:conv1dHooke} shows that convergence is achieved for all \revision{EEXX+PT2} excitation energies of the 1D Hooke's atom with only a few KS orbitals. Higher-energy orbitals have a small impact on excitation energies as the KS energy differences become large. Due to the near-degeneracies between KS states $(2,2)$ and $(1,3)$ and between $(2,3)$ and $(1,4)$, the changes of the $I=2$ and $I=4$ excitation energies are big when orbitals 3 and 4 are included, respectively. The excitation energies of the 1D charge-transfer box and 1D flat box (see supplemental material\cite{supplemental}) also converge quickly with respect to the number of orbitals, since the orbital energies of these systems increase rapidly.

\begin{figure}[htbp]
\includegraphics[width=\columnwidth]{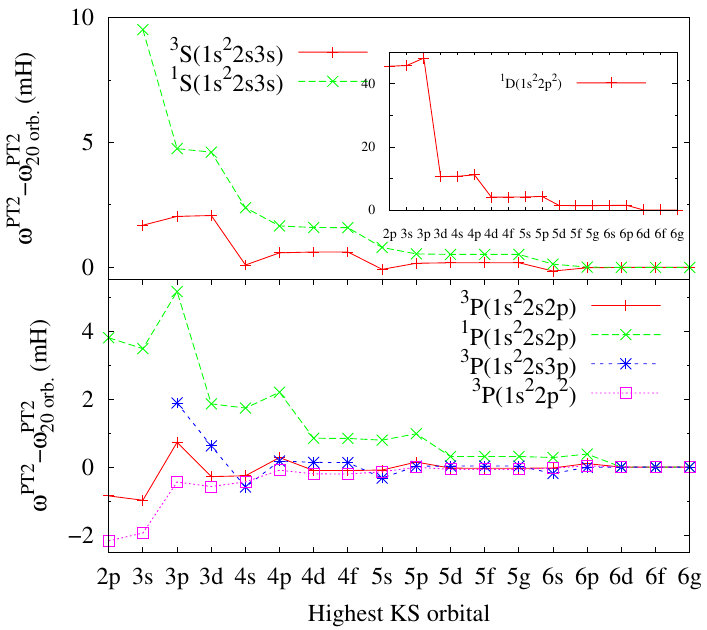}
\caption{Convergence of \revision{EEXX+PT2} excitation energies of the Be atom with respect to the number of KS \revision{orbitals}. Results for S, P and D states are shown in the upper panel, lower panel and the inset respectively. The excitation energies calculated with 20 KS orbitals are used as references. The horizontal axis is labeled by the highest orbital. The differences between the last two points of each curve are all smaller than $10^{-4}$ mH.}
\label{fig:results:discussion:conv3dBe}
\end{figure}

Fig. \ref{fig:results:discussion:conv3dBe} plots the \revision{EEXX+PT2} excitation energies of the Be atom with different numbers of KS orbitals included. The He atom have similar trends (see supplemental material\cite{supplemental}). Comparing with 1D cases, higher-energy orbitals can affect the excitation energies significantly since the orbital energies form a Rydberg series. More orbitals are needed to achieve convergence in 3D atoms.

The excitation energies in Fig. \ref{fig:results:discussion:conv3dBe} only change notably when orbitals with certain angular momentum quantum numbers are included, since otherwise many of the newly included matrix elements in Eq. \parref{eqn:PT2:derivPT2work} would vanish due to symmetry. Therefore one can achieve convergence with fewer orbitals for atoms by ignoring the orbitals whose angular momentum quantum numbers differ too much from those of the orbitals in the KS ensemble. As an example, the \revision{EEXX+PT2} excitation energy of the $^1$S(1s$^2$2s3s) state \revision{of Be} calculated with all orbitals (upto 6g) only differ from that calculated with only s and p orbitals (upto 6p) by 0.007 mH.

\section{Conclusion}
\label{sec:conclusion}
We apply GLPT on EDFT and derive the orbital-dependent ensemble PT2 correlation energy functional. We test the performance of the ensemble PT2 functional on 1D model systems and 3D atoms using the DEC method in our previous work. For 1D model systems, inclusion of $E^\txtPT2_{\txtc,\wt}$ improves the exchange-only \revision{EEXX} results in general, but the errors of \revision{EEXX+PT2} are at the same orders of magnitude as those of \revision{EEXX}. We show that the ensemble correlation effect on excitation energies of an orbital-dependent functional consists of contributions from the $\wt$-dependence of $E_{\txtc,\wt}$ and from $v\c$. The ratio of these two can differ a lot for different systems, and double excitation and charge-transfer excitation represent two extremes, \revision{which} are dominated by the $\wt$-dependence of $E_{\txtc,\wt}$ and by $v\c$ respectively. Calculation results suggest that PT2 may not be enough to properly approximate the ensemble correlation effect, which may need higher order terms.

For 3D He and Be atoms, the \revision{EEXX+PT2} results are generally worse than \revision{EEXX}. The failure of \revision{EEXX+PT2} is probably related to the inconsistency between the $E^\txtPT2_{\txtc,\wt}$ functional and the exact $v\c$ used in the calculation. This problem only affects orbital-dependent functionals, and the OEP of $E^\txtPT2_{\txtc,\wt}$ is needed to completely assess the performance of the ensemble PT2 correlation. However, we find the ensemble PT2 OEP more difficult to evaluate than in ground-state due to ensemble sums and divergences, and it is beyond the scope of this paper to develop proper approximations for it. We also find that the commonly used ground-state approximation of neglecting singly-excited matrix elements is applicable in EDFT with slightly larger error. Because of the problems associated with the ensemble PT2, a double hybrid functional mixing in a part of PT2 in the generalized EKS scheme\cite{GK21} might be a more viable approach for developing approximated ensemble Hxc energy functionals.

$E^\txtPT2_{\txtc,\wt}$ contains sums over all EKS Slater determinants, and we studied the convergence of \revision{EEXX+PT2} excitation energies with respect to the number of KS orbitals. We find that in 1D systems convergence is achieved with only one or two extra orbitals than the minimal case, and the excitation energies only change slightly when higher-energy orbitals are included. Although more orbitals are needed for convergence of 3D atoms due to the orbital energies resemble the Rydberg series, convergence is still achieved fairly fast, allowing the use of $E^\txtPT2_{\txtc,\wt}$ in practical calculations.

\section*{Acknowledgements}
This work is supported by the National Natural Science Foundation of China Grant No. 11804314.

\appendix
\section{\revision{Details in GLPT for EDFT}}
\label{app:GLPT}
\revision{
The scaled Hamiltonian $\hat{H}_{\lambda,\bracew}$ is equal to $\hat{H}^{(0)}_\bracew+\lambda\hat{H}'_{\lambda,\bracew}$ in GLPT, and the zeroth-order and perturbative Hamiltonians are
\ben
\hat{H}^{(0)}_\bracew=\hat{H}_{\txts,\bracew}=\hat{T}+\hat{V}_{\txts,\bracew},
\label{eqn:theory:GLPT:H0}
\een
and
\ben
\begin{split}
\lambda\hat{H}'&=\lambda\hat{V}\ee+\hat{V}_{\lambda,\bracew}-\hat{V}_{\txts,\bracew}\\
&=\lambda(\hat{V}\ee+\hat{V}^{(1)}_\bracew)+\sum_{p=2}^\infty\lambda^p\hat{V}^{(p)}_\bracew.
\end{split}
\label{eqn:couplingconst:pertH}
\een
Eq. \parref{eqn:couplingconst:pertH} differs from the Rayleigh-Schr\"{o}dinger perturbation theory since the perturbation is dependent on $\lambda$.

GLPT expands $v_{\lambda,\bracew}(\br_i)$ at $\lambda=0$ as
\ben
v_{\lambda,\bracew}(\br)=\sum_{p=0}^\infty \lambda^p v_\bracew^{(p)}(\br),
\label{eqn:couplingconst:vseries}
\een
and $E_{\lambda,\bracew}$ and $\Phi_{ik,\lambda,\bracew}$ can be expanded similarly. Since $v_{0,\bracew}(\br)=v_{\txts,\bracew}(\br)$ and $v_{1,\bracew}(\br)=v\ext(\br)$, we can derive the expression for the ensemble Hxc potential from Eqs. \parref{eqn:theory:background:vdecomp} and \parref{eqn:couplingconst:vseries} as
\ben
v_{\txtHxc,\bracew}(\br)=-\sum_{p=1}^\infty v^{(p)}_\bracew(\br).
\label{eqn:theory:GLPT:vHxc}
\een
For the ensemble energy, we have $E_{0,\bracew}=E^\text{KS}_\bracew=T_{\txts,\bracew}+\int\intd^3r\,n_\bracew(\br)[v\ext(\br)+v_{\txtHxc,\bracew}(\br)]$ and $E_{1,\bracew}=E_\bracew$. The ensemble Hxc energy is derived similarly as in Eq. \parref{eqn:theory:GLPT:vHxc}:
\ben
E_{\txtHxc,\bracew}=\sum_{p=1}^\infty E^{(p)}_\bracew -\int\intd^3r\,n_\bracew(\br)\sum_{p=1}^\infty v^{(p)}_\bracew(\br).
\label{eqn:theory:GLPT:EHxc}
\een

Inserting the $\lambda$-expansions into Eq. \parref{eqn:couplingconst:schrodingereqn} and collecting different orders of $\lambda$, the equations from $\lambda^0$ to $\lambda^2$ are
\begin{align}
\lambda^0:&\hat{H}^{(0)}_\bracew\ket{\Phi^{(0)}_{ik,\bracew}}=E^{(0)}_{i,\bracew}\ket{\Phi^{(0)}_{ik,\bracew}}\label{eqn:theory:GLPT:schrodinger0th}\\
\lambda^1:&\hat{H}^{(0)}_\bracew\ket{\Phi^{(1)}_{ik,\bracew}}+(\hat{V}\ee+\hat{V}^{(1)}_\bracew)\ket{\Phi^{(0)}_{ik,\bracew}}\nonumber\\
&=E^{(0)}_{i,\bracew}\ket{\Phi^{(1)}_{ik,\bracew}}+E^{(1)}_{i,\bracew}\ket{\Phi^{(0)}_{ik,\bracew}}\label{eqn:theory:GLPT:schrodinger1st}\\
\lambda^2:&\hat{H}^{(0)}_\bracew\ket{\Phi^{(2)}_{ik,\bracew}}+(\hat{V}\ee+\hat{V}^{(1)}_\bracew)\ket{\Phi^{(1)}_{ik,\bracew}}+\hat{V}^{(2)}_\bracew\ket{\Phi^{(0)}_{ik,\bracew}}\nonumber\\
&=E^{(0)}_{i,\bracew}\ket{\Phi^{(2)}_{ik,\bracew}}+E^{(1)}_{i,\bracew}\ket{\Phi^{(1)}_{ik,\bracew}}+E^{(2)}_{i,\bracew}\ket{\Phi^{(0)}_{ik,\bracew}},
\label{eqn:theory:GLPT:schrodinger2nd}
\end{align}
where $\hat{H}^{(0)}_\bracew=\hat{T}+\hat{V}_{\txts,\bracew}$ and $\ket{\Phi^{(0)}_{ik,\bracew}}=\ket{\Phi_{ik,\bracew}}$.

The first and second order corrections to the EKS energy can be derived from Eq. \parref{eqn:theory:GLPT:schrodinger2nd} as
\ben
E^{(1)}_\bracew=E_{\txtHx,\bracew}+\int\intd^3r\,v^{(1)}_\bracew(\br)n_\bracew(\br),
\label{eqn:theory:GLPT:E1}
\een
and
\ben
\begin{split}
E^{(2)}_\bracew=&\sum_{i=0}^I w_i\sum_{k=1}^{g_i}\matelem{\Phi_{ik,\bracew}}{\hat{V}\ee+\hat{V}^{(1)}_\bracew}{\Phi^{(1)}_{ik,\bracew}}\\
&+\int\intd^3r\,v^{(2)}_\bracew(\br)n_\bracew(\br).
\end{split}
\label{eqn:theory:GLPT:E2}
\een
Eq. \parref{eqn:theory:GLPT:E1} indicates that $v^{(1)}_\bracew(\br)=-v_{\txtHx,\bracew}(\br)=-\delta E_{\txtHx,\bracew}/\delta n(\br)$ since $\delta E^{(1)}_\bracew/\delta n(\br)=0$ due to variational principle. Inserting Eq. \parref{eqn:theory:GLPT:E1} and \parref{eqn:theory:GLPT:E2} into the ensemble energy decomposition Eq. \parref{eqn:theory:GLPT:enseng}, and noticing that
\ben
E^{(0)}_\bracew=T_{\txts,\bracew}+\int\intd^3r\,n_\bracew(\br)\left[v\ext(\br)-\sum_{p=1}^\infty v^{(p)}_\bracew(\br)\right],
\label{eqn:theory:GLPT:E0}
\een
we arrive at the formula for the ensemble PT2 correlation energy Eq. \parref{eqn:theory:GLPT:EcPT2}.

\section{Justification of neglecting the fixed-orbital $\wt$-derivative of $v_{\Hx,\wt}$ in DEC/PT2}
\label{sec:OEP:justification}
The second term of Eq. \parref{eqn:PT2:derivPT2work} is neglected the calculations, and we provide a justification here. Since the DEC method is evaluated at $\wt=0$, this term reduces to
\begin{multline}
\sum_{\tilde{j}\ne0}^\infty\sum_{q=1}^{\tilde{g}_{\tilde{j}}}\frac{1}{\calE^\txtKS_0-\calE^\txtKS_j}\Bigg\{\matelem{\tilde{\Phi}_{\tilde{j}q}}{\hat{V}\ee-\hat{V}\Hx}{\tilde{\Phi}_0}^*\\
\times\matelem{\tilde{\Phi}_{\tilde{j}q}}{-\left.\frac{\partial\hat{V}_{\txtHx,\wt}[n_{\txts,\wt}[\{\phi_\mu\}]]}{\partial \wt}\right|_{\phi_\mu=\phi^\txtKS_\mu}}{\tilde{\Phi}_0}+\text{c.c.}\Bigg\},
\label{eqn:OEP:justification:secondterm}
\end{multline}
where we assume that the ground state is non-degenerate for simplicity.

$v_{\Hx,\wt}$ is a one-body potential, so is its fixed-orbital $\wt$-derivative. Therefore the summations over $\tilde{j}\ne0$ and $q$ in Eq. \parref{eqn:OEP:justification:secondterm} only involve `singly-excited' $\tilde{\Phi}_{\tilde{j}q}$ that differs from $\tilde{\Phi}_0$ by one orbital according to the Slater-Condon rules. The $\hat{V}\ee-\hat{V}\Hx$ in Eq. \parref{eqn:OEP:justification:secondterm} is $\hat{H}-\hat{H}^\txtKS$ truncated at the first order. We have $\matelem{\tilde{\Phi}_{\tilde{j}q}}{\hat{H}^\txtKS}{\tilde{\Phi}_0}=0$ since $\tilde{j}\ne0$ , so the magnitude of $\matelem{\tilde{\Phi}_{\tilde{j}q}}{\hat{V}\ee-\hat{V}\Hx}{\tilde{\Phi}_0}$ is approximately that of $\matelem{\tilde{\Phi}_{\tilde{j}q}}{\hat{H}}{\tilde{\Phi}_0}$, which vanishes for Hartree-Fock (HF) single-Slater-determinant wavefunctions according to Brillouin's theorem. Although this matrix element does not vanish for Kohn-Sham (KS) Slater determinants, it is small in most cases since the KS Slater determinants resemble the HF ones\cite{SH99}. Therefore the second term of Eq. \parref{eqn:PT2:derivPT2work} can be neglected in EEXX+PT2 calculations in most cases.

We carry out EEXX+PT2 calculations with this term for the Be atom to demonstrate the validity of this approximation. The fixed-orbital $\wt$-derivative of $v_{\txtHx,\wt}$ term is calculated with the Slater approximation (see supplemental material\cite{supplemental}). These calculations must use the full ensemble containing all the states from the ground state to the interested excited multiplet, so the computational cost is much larger than the bi-ensemble calculations in the main text. Table \ref{table:OEP:Be:result} lists the differences of excitation energies with and without this term, and these differences are at least two orders of magnitudes smaller than the error in EEXX+PT2 shown in the main text. This demonstrates that the approximation of neglecting this term is viable.

\begin{table}[htbp]
\revision{
\begin{tabular}{cc}
\hline\hline
I & $\omega_I^\text{full}-\omega_I^\text{part}$ (mH)\\
\hline
$^3$P(1s$^2$2s2p) & 0.02396 \\
$^1$P(1s$^2$2s2p) & 0.03065 \\
$^3$S(1s$^2$2s3s) & 0.1580 \\
$^1$S(1s$^2$2s3s) & 0.1310 \\
$^1$D(1s$^2$2p$^2$) & 0.04902 \\
$^3$P(1s$^2$2s3p) & 0.1510 \\
$^3$P(1s$^2$2p$^2$) & 0.03527 \\
\hline\hline
\end{tabular}
}
\caption{\revision{Difference between the excitation energies of Be calculated by EEXX+PT2 with and without the fixed-orbital $\wt$-derivative of $v_{\txtHx,\wt}$ term, which are denoted by $\omega_I^\text{full}$ and $\omega_I^\text{part}$ respectively. The calculation parameters are the same as in the main text.}}
\label{table:OEP:Be:result}
\end{table}

\section{1D flat box}
\label{app:1dbox}
We carry out DEC and TDDFT calculations of a two-electron 1D flat box whose external potential is
\ben
v\ext(x)=\left\{\begin{array}{ll}
\infty & x\in(-\infty,0]\cup[1,\infty),\\
0 & x\in(0,1),
\end{array}\right.
\label{eqn:results:1D:box:vext}
\een
and the soft-Coulomb interaction Eq. \parref{eqn:results:1D:veesoftcoulomb}. We use $\alpha=0.01$ to ensure that the differences between exact and KS excitation energies are large, so that the improvement of EEXX+PT2 results over EEXX can be seen clearly. We use a grid with 1001 points with grid-point spacing 0.001 for $x\in[0,1]$. Table \ref{table:1dbox:result} lists the calculation results.

\begin{table}[htbp]
\revision{
\footnotesize
\begin{tabular}{cccccccc}
\hline\hline
 & & & \multicolumn{5}{c}{$\Delta\omega_I=\omega_I^\text{DEC}-\omega_I^\text{exact}$ (mH)}\\
\cline{4-8}
$I$ & $\omega_I^\text{exact}$ (H) & $\omega_I^\txtKS$ (H) & EEXX & +$E\c^\txtPT2$ & +$v\c$ & +PT2 & $\substack{\text{+PT2}\\\text{(no single)}}$\\
\hline
\multicolumn{8}{c}{Singlet}\\
\hline
2(1,2) & 15.62 & 13.88 & -78.40 & 28.41 & 31.76 & 138.6 & 104.2\\
3(2,2) & 28.86 & 27.76 & -145.2 & -220.0 & 75.16 & 0.3752 & 17.04\\
5(1,3) & 39.93 & 38.60 & -302.0 & 13.59 & -261.9 & 53.66 & 51.76\\
7(2,3) & 54.49 & 52.48 & -153.9 & -212.9 & -3.650 & -62.67 & -40.72\\
9(1,4) & 74.05 & 73.12 & -281.3 & -32.25 & -236.9 & 12.21 & 20.92\\
10(3,3) & 77.93 & 77.20 & -18.99 & -107.4 & 61.15 & -27.21 & -38.43\\
\hline
\multicolumn{8}{c}{Triplet}\\
\hline
1(1,2) & 12.44 & 13.88 & -219.7 & -144.7 & -109.5 & -34.57 & -2.608\\
4(1,3) & 37.70 & 38.60 & -132.9 & -42.36 & -92.82 & -2.292 & 8.062\\
6(2,3) & 52.08 & 52.48 & -246.8 & -123.0 & -96.53 & 27.20 & 18.81\\
8(1,4) & 72.61 & 73.12 & -136.3 & -34.64 & -91.81 & 9.829 & 20.89\\
\hline\hline
\end{tabular}
}
\caption{\revision{Errors in the excitation energies of the 1D flat box. 6 orbitals are used in EEXX+PT2 calculations. The occupied KS orbitals are shown in parentheses after $I$, and the ground state is $(1,1)$. 7 orbitals are used in the calculations.}}
\label{table:1dbox:result}
\end{table}

}

\bibliographystyle{unsrt}
\bibliography{PT2}

\end{document}